\documentclass[prb,10pt,twocolumn,superscriptaddress]{revtex4}
\usepackage{amsmath}
\usepackage{amssymb}
\usepackage{dsfont}
\usepackage{graphicx}
\usepackage{verbatim}
\usepackage{color}
\usepackage{subfigure}
\usepackage{hyperref}
\newcommand{\cop}[2]{c_{#1}^{#2}} 
\newcommand{\cd}[1]{c_{#1\downarrow}^+} 
\newcommand{\cu}[1]{c_{#1\uparrow}^+} 
\newcommand{\au}[1]{c_{#1\uparrow}^{}} 
\newcommand{\ket}[1]{|#1\rangle} 
\newcommand{\bra}[1]{\langle#1|} 
\newcommand{\Fig}[1]{Fig.\ref{#1}} 
\newcommand{\eq}[1]{Eq.\ref{#1}} 
\newcommand{\neqn}[1]{\begin{equation}#1\end{equation}}

\begin{document}


\title{Electron-Electron interactions, topological phase and optical properties of a charged 
artificial benzene ring} 
\author{Isil Ozfidan}
\address{Physics Department, University of Ottawa, Ottawa, Canada}
\author{Milos Vladisavljevic} 
\address{Physics Department, University of Ottawa, Ottawa, Canada}
\author{Marek Korkusinski}
\address{Emerging Technologies Division, 
National Research Council of Canada Ottawa, K1A OR6, Canada}
\author{Pawel Hawrylak}
\address{Physics Department, University of Ottawa, Ottawa, Canada}

\begin{abstract}
We present a theory of the electronic and optical properties of a charged artificial benzene ring (ABR).
The ABR is described by the extended Hubbard model solved
using exact diagonalization methods  in both real and Fourier space as a function of tunneling matrix  element t, Hubbard on-site repulsion U and 
inter-dot interaction V. In the strongly interacting case we present exact analytical results for the spectrum of the hole in a half-filled ABR dressed 
by spin excitations of remaining electrons.  The spectrum is  interpreted in terms of
the appearance of a topological phase associated with an effective gauge field piercing through the ring. 
We show that the maximally  spin polarized (S=5/2) and maximally spin depolarised (S=1/2) states are the lowest energy, orbitally non degenerate, states. 
We discuss the evolution of the phase diagram and level crossings as interactions are switched
off and the ground state becomes  spin non-degenerate  but orbitally degenerate S=1/2.
We present a theory of optical absorption spectra and show that the  evolution of the ground and excited states, level crossings
and presence of artificial gauge can be detected optically.

\end{abstract}

\maketitle
\section{Introduction}
There is currently interest in developing controlled quantum many-body systems using semiconductor quantum dots and molecules as means to understand the many-body problem as well as for applications in nanoelectronics, nanospintronics, and quantum information processing.
Single \citep{hawrylak_gould_prb1999,ciorga_sachrajda_prb2000}, 
double \citep{elzerman_hanson_nat2004,petta_johnson_sci2005,pioro_abolfath_prb2005}, 
triple \citep{gaudreau_studenikin_prl2006,schroer_greentree_prb2007,ihn_sigrist_njp2007,rogge_haug_prb2008,laird_taylor_prb2010,amaha_hatano_prb2012} 
and
 quadruple lateral gated quantum dot molecules in GaAlAs/GaAs heterojunctions or with dangling bonds on silicon surface have been demonstrated experimentally  \citep{shulman_dial_science2012,thalineau_hermelin_apl2012,haider_pitters_prl2009} 
and extensively studied theoretically \citep{hsieh_shim_ropp2012,hawrylak_korkusinski_ssc2005,korkusinski_gimenez_prb2007,gimenez_korkusinski_prb2007,delgado_shim_prl2008,gimenez_hsieh_prb2009,hawrylak_prl1993,brum_hawrylak_sm1997,lobos_aligia_prl2008,nisikawa_oguri_prb2006, sharma_hawrylak_prb2011,scarola_dassarma_pra2005,mizel_lidar_prb2004, ozfidan_ssc2013,castelano_prb2006,vlad_thesis_2014}.  
The capability to localize electrons in artificial lateral quantum dot molecules opens up the possibility of exploring the properties of the 1D Hubbard model, a model of strongly correlated electrons \citep{Nagaoka,caspers_iske_physe1989, penc_hallberg_prb1997,lieb_wu_prl1990,ogata_shiba_prb1990,mila_penc_jes2000,dagotto_prl2008}.  The 1D Hubbard model of benzene rings is of  recent interest in the context of charge separation in mesoscopic rings \citep{hallberg_aligia_prl2004,aligia_hallberg_prl2004,rincon_aligia_prb2009}, optical properties of strongly correlated oxides \citep{dagotto_prl2008}, quantum tunnelling in vertically coupled rings\citep{castelano_prb2006}, inelastic co-tunnelling in double, triple, and benzene like quantum dot molecules\citep{begemann_prb2010}, 
electron localization\citep{ballester_jap2012}, transport\citep{begemann_prb2008,tosi_jpcm2012}, quantum interference\citep{ke_nl2008} and Coulomb blockade\citep{hettler_prl2003}. 
Of particular interest here are the properties of charged rings where the orbitally degenerate ground state  leads to non-Fermi liquid behaviour and Kondo effect in transport\citep{begemann_prb2008}.
Understanding artificial benzene rings is also important for the understanding of graphene. There have been several experimental realizations of artificial graphene as a platform to study Dirac fermions and topological phases \citep{simoni_apl2010, singha_science2011, gomes_nature2012}.

\begin{figure}[th]
\begin{center}
\includegraphics[width=0.7\linewidth]{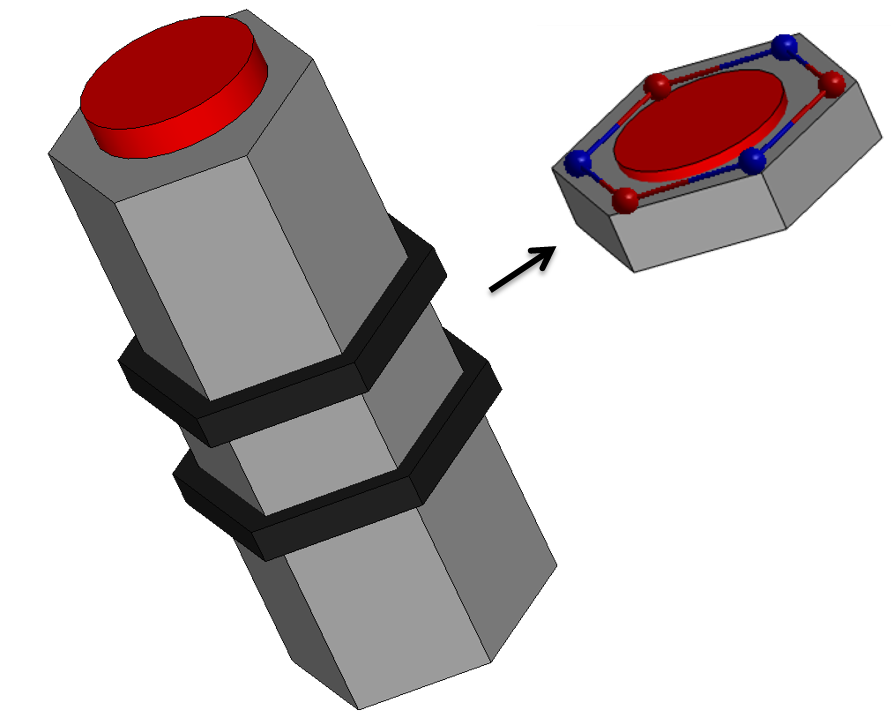}
\end{center}
\caption{(a) Confining 2D electron gas in a nanowire. Proposed model for realization of an artificial benzene ring by confining electrons to 6 corners of a hexagonal nanowire}
\label{fig1}
\end{figure}

The artificial benzene rings could be now realized in  hexagonal semiconductor nanowires. Ballester et al.\citep{ballester_jap2012} investigated theoretically a quasi-2D hexagonal nanostructure cut out of an AlAs/GaAs/AlAs core shell nanowire. They have shown that in such hexagonal ring, the states would weakly localize at the 6 corners. However, the weak localization resulted in features different from what is expected in a benzene ring, for example the five electron ground state was not doubly degenerate.
Recently Funk et al. \citep{funk_abstreiter_nl2013} demonstrated confinement of electrons in 6 one dimensional electron channels localized to the 6 corners of the hexagonal core shell nanowire. If one was to fabricate a wrap-around gate, shown in \Fig{fig1}, one could create 6 quantum dots, one in each 1D channel. Additional gates, not shown, could control the tunnelling of electrons between different dots. Tunnelling between the dots and Coulomb interactions could be tuned by changing the size of the structure to modify the distance between dots. After construction, the interactions could be altered by changing the gate potentials or adding electrodes to individually control the interactions among specific dots.\citep{ballester_jap2012, hsieh_shim_ropp2012,blomers_nt2013, albert_jcg2014, poole_nt2012, polini_nn2013, vlad_thesis_2014} The advantage of such a system would be its tunability in comparison to natural benzene. 

Motivated by the experiments and theoretical interest, we provide here a theory of the electronic and optical properties of an artificial, charged benzene ring (ABR) molecule described by the Hubbard model with tunable parameters; inter-dot tunnelling $t$ and Coulomb interactions $U$ and $V$. 
In the strongly interacting case, we present exact analytical results for the spectrum of the hole in a half-filled ABR dressed by spin excitations of remaining electrons. The spectrum is interpreted in terms of
the appearance of a topological phase associated with an effective gauge field piercing through the ring. We show that the maximally spin polarized (S=5/2) and maximally spin depolarised (S=1/2) states are the lowest energy, orbitally non degenerate states. We discuss the evolution of the phase diagram and level crossings as interactions are switched
off and the ground state becomes spin non-degenerate, S=1/2, but orbitally degenerate. We present a theory of the optical absorption spectra and show that the evolution of the ground and excited states, level crossings
and artificial gauge can be detected optically.
\section{The Model}

Following previous work \citep{ozfidan_ssc2013}, the artificial benzene molecule \citep{cooper_nat1986, pauling_nat1987, messmer_nat1987, harcourt_nat1987} is assumed to have one spin-degenerate orbital per quantum dot, with the  molecule containing up to $N_e=12$ electrons \citep{hsieh_shim_ropp2012,gimenez_korkusinski_prb2007}. Its electronic properties are described microscopically within the extended Hubbard model\citep{hawrylak_korkusinski_ssc2005} which, in the real-space basis, is given as\citep{ozfidan_ssc2013}:
\begin{eqnarray}
 \hat{H}=\sum_{\sigma,i=1}^{6}E_i \cop{i\sigma}{+}\cop{i\sigma}{}-\sum_{\sigma,\langle i,j \rangle}^{6}t_{ij}\cop{i\sigma}{+}\cop{j\sigma}{} \nonumber\\ +\sum_{i=1}^{6}U_i
n_{i\downarrow}n_{i\uparrow}+\frac{1}{2}\sum_{\langle i,j \rangle}V_{ij}\varrho_i\varrho_j.
\label{ham}
\end{eqnarray}
Here $\cop{i\sigma}{+}(\cop{i\sigma}{})$ are operators creating(annihilating) a spin-$\sigma$ electron on a localized quantum dot orbital $i$ with energy $E_i$, while the spin and charge density are expressed as $n_{i\sigma}=\cop{i\sigma}{+}\cop{i\sigma}{}$ and
$\varrho_i=n_{i\downarrow}+n_{i\uparrow}$, respectively.  The on-site interaction between two electrons on each dot is given by $U_i$ while $t_{ij}$ and $V_{ij}$ characterize the tunnelling and Coulomb matrix elements between dots $i$ and $j$. We only retain the nearest-neighbour (NN), $\langle ij\rangle$, elements of both. 
The Hamiltonian matrix for a single electron tunnelling between 6 dots is then explicitly written as
\begin{equation}
\hat{H}=\left[\begin{array}{cccccc}
0&t&0&0&0&\tau \\
t&0&t&0&0&0 \\
0&t&0&t&0&0 \\
0&0&t&0&t&0 \\
0&0&0&t&0&t \\
\tau&0&0&0&t&0
\end{array}\right],
\label{Hammat}
\end{equation} 
where $\tau$ represents the tunnelling between dot 1 and dot 6. $\tau=t$ for the closed ring and it is $\tau=0$ for a finite chain.  
  
\begin{figure}[h]
\begin{center}
\includegraphics[width=0.98\linewidth]{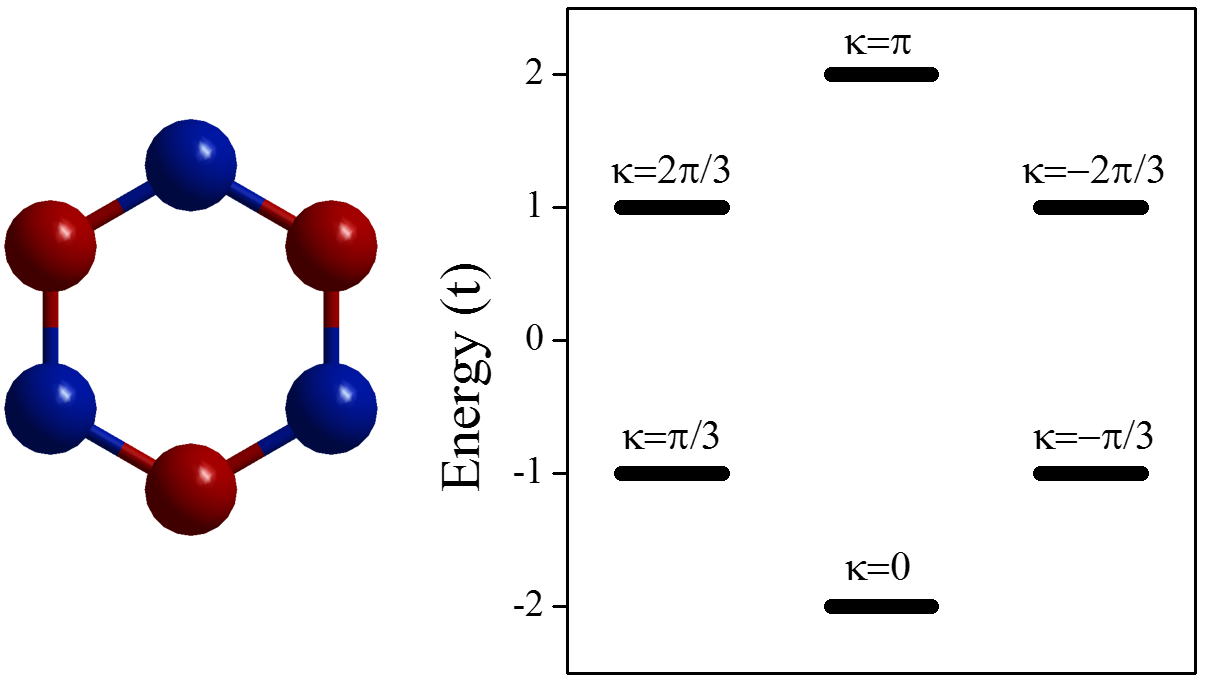}
\end{center}
\caption{(a) The ABR structure (b) one-electron spectrum labelled with wavevectors}
\label{fig2}
\end{figure}

The Hamiltonian \eq{ham}, can also be rotated into the Fourier space of itinerant electrons using a Fourier transform of the real-space creation/annihilation operators,
\begin{equation}
a_{\kappa_i}^+=\frac{1}{\sqrt{6}}\sum_{j=1}^6 e^{i \kappa_i(j-1)}\cop{j}{+},
\label{forop}
\end{equation}
where
$\kappa_i=\{0,\pm \pi/3, \pm2\pi/3, \pi\}$ are the 6 allowed wave-vectors. The  operators $a_{\kappa_i}^+ (a_{\kappa_i})$ create(annihilate) an electron on a Fourier state $\ket{\kappa_i}$.
Assuming all dots  on resonance, i.e., $E_i=E$, $U_i=U$, $V_{ij}=V$, the rotated Hamiltonian becomes 
\begin{eqnarray}
\hat{H}&=&\sum_{\sigma,i}^{6}\epsilon_{\kappa_i} a_{\kappa_i\sigma}^+a_{\kappa_i\sigma}^{} \nonumber\\
&&+\frac{1}{2}\sum_{ijkl\sigma\sigma'}\bra{\kappa_i\kappa_j}\mathcal{V}_{ee}\ket{\kappa_k\kappa_l}a_{\kappa_i\sigma}^+a_{\kappa_j\sigma'}^+a_{\kappa_k\sigma'}^{}a_{\kappa_l\sigma}^{}.
\label{forcol}
\end{eqnarray}
The transformation into itinerant molecular $\ket{\kappa_i}$ states diagonalizes the Hamiltonian with the following eigenvalues $\epsilon_{\kappa_i}=E+2t\cos{\kappa_i}$ giving the single particle spectrum shown in \Fig{fig2}.

The second term in \eq{forcol} describes the Coulomb interaction matrix elements between molecular states,
\begin{equation}
\bra{\kappa_i\kappa_j}\mathcal{V}_{ee}\ket{\kappa_k\kappa_l}=\frac{U+2V\cos{(\kappa_l-\kappa_i)}}{6}\delta(\kappa_i+\kappa_j,\kappa_k+\kappa_l).
\label{coulmat}
\end{equation}
We note that the total wavevector $\kappa_{tot}=\sum_i^{N_e}\kappa_i$, is conserved in Coulomb scattering. 

With the Hamiltonian established, we expand the many-body states in $N_e$-electron configurations, $\ket{\alpha}$, created by distributing $N_e$ electrons on 6 molecular orbitals obeying the Pauli exclusion principle, where $\ket{\alpha}=\prod_{i=1,N_e} \cop{i\sigma}{+}\ket{0}$
and $\ket{0}$ is the vacuum. Similarly, we construct the many-electron states using the real space orbitals. By constructing a real-space or a Fourier-space Hamiltonian matrix for $N_e$ electrons with spin $S_z$, and diagonalizing the matrix, we obtain the corresponding eigenenergies $E_{f_{N_e}}$ and eigenvectors $| f_{N_e}\rangle$ in terms of real or Fourier space orbitals. 

The optical properties of the ABR are described using the Fermi's golden rule \citep{ozfidan_prb2014}. The transition rate from the ground state to excited states of the $N_e$ electron system via absorption of a photon with energy $\omega$ is given by:
\begin{equation}
A_{N_e}(\omega) = \sum_{f}W_{GS}|\langle f_{N_e} | \hat{P}^+ | GS_{N_e} \rangle |^2 \delta(E_{f_{N_e}}-E_{GS}-\omega),
\end{equation}
where $|GS_{N_e}\rangle$ is the $N_e$ electron ground state with energy $E_{GS}$, $W_{GS}$ is the probability of its occupation, and $\ket{f_{N_e}}$ is the excited state with energy $E_{f_{N_e}}$. The polarization operator $P^+$ moves an electron from a filled state to a higher energy, unoccupied state, while annihilating a photon, $\hat{P}^+=\sum_{\kappa_j,\kappa_i,\sigma} d(\kappa_j,\kappa_i) a_{\kappa_j\sigma}^+ a_{\kappa_i\sigma}^{}$ \citep{ozfidan_prb2014}. 

Following our previous work\citep{ozfidan_prb2014}, the dipole element $d(\kappa_j,\kappa_i)$ expressed in the basis of localized orbitals is written as :
\begin{equation}
\bra{l} \vec{\varepsilon}\cdot\vec{r} \ket{j}
= D \vec{\varepsilon}\cdot (\vec{R}_j-\vec{R}_l)\delta_{\langle
  lj\rangle}+\vec{\varepsilon}\cdot \vec{R}_l\delta_{lj},
 \label{eqdip}
\end{equation} 
where $D$ is the dipole strength coefficient (in units of inter dot distance) calculated for NN elements using basis orbitals and $\vec{\varepsilon}$ is the polarization of light. In what follows we will use the numerical values obtained for graphene $p_z$ orbitals \citep{ozfidan_prb2014}.

\begin{figure}[ht]
\begin{center}
\includegraphics[scale=0.3]{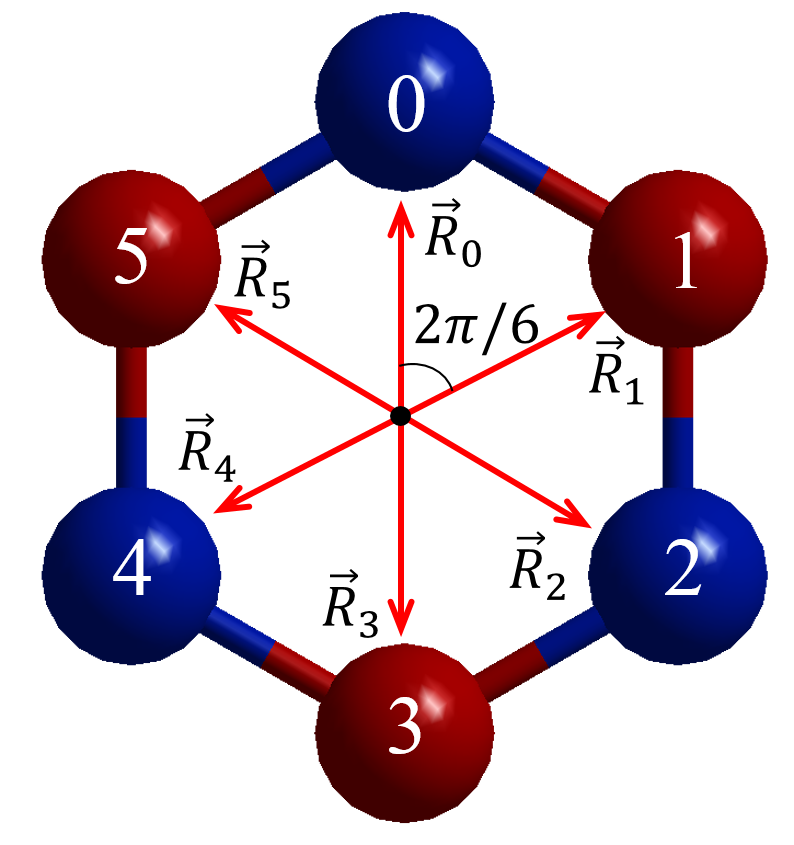}
\end{center}
\caption{The six dipole moments  measured from the center of a benzene ring.} 
\label{fig3}
\end{figure}

Due to the hexagonal structure of the ABR, the vectors extending from the center of the ABR to the localized orbitals are equal in magnitude, with directions varying as multiples of $2\pi/6$, as depicted in \Fig{fig3}. As a result, the dot product between the polarization of light and the vector $\vec{R}_{m}$ pointing from the center of the ring to each localized orbital $m$  that appears in \eq{eqdip} can be simplified as 
\neqn{\varepsilon_{\pm} \cdot \vec{R}_{m}=|R|e^{\pm i m2\pi/6}. \label{dotp}}
The dipole elements between molecular states are calculated by writing them out explicitly in terms of the atomic orbitals. For light that is circularly polarized, $\varepsilon_\pm$, the dipole element between molecular states can be expanded as
\begin{equation}
\bra{\kappa_j}\varepsilon_\pm\cdot\vec{r}\ket{\kappa_i}=\frac{1}{6}\sum_{ p,q}^6 e^{i\left(\kappa_iq-\kappa_jp\right)}\bra{p}\varepsilon_\pm\cdot\vec{r}\ket{q},
\end{equation}
where $p,q$ are localized $p_z$ orbitals. We can open up the sum using \eq{eqdip} and retain $\delta_{pq}$ and $\delta_{\langle pq\rangle}$ elements since we are only including up to nearest neighbour tunnelling. Then the expression above becomes
\begin{eqnarray}
\bra{\kappa_j}\varepsilon_\pm\cdot\vec{r}\ket{\kappa_i}&=&\frac{1}{6}\sum_p
[e^{i\left(\kappa_i(p-1)-\kappa_jp\right)}\varepsilon_\pm\cdot D(\vec{R}_{p-1}-\vec{R}_{p}) \nonumber \\
&&+e^{i\left(\kappa_ip-\kappa_jp\right)}\varepsilon_\pm\cdot\vec{R}_p \nonumber \\
&&+e^{i\left(\kappa_i(p+1)-\kappa_jp\right)}\varepsilon_\pm\cdot D(\vec{R}_{p+1}-\vec{R}_{p})]. \nonumber \\
\end{eqnarray}
Collecting $R_i$ and using \eq{dotp} for the dot products, we obtain
\begin{eqnarray}
\bra{\kappa_j}\varepsilon_\pm\cdot\vec{r}\ket{\kappa_i}&=&\frac{|R|}{6}\sum_p[e^{ip\left(\kappa_i-\kappa_j\pm \pi/3\right)}\left(1-2D\cos(\kappa_i)\right) \nonumber \\
&&+D e^{ip\left(\kappa_i-\kappa_j\pm \pi/3\right)}e^{-i(\kappa_i\pm \pi/3)} \nonumber \\
&&+D e^{ip\left(\kappa_i-\kappa_j\pm \pi/3\right)}e^{+i(\kappa_i\pm \pi/3)}],
\end{eqnarray}
which, once simplified gives,
\begin{eqnarray}
\bra{\kappa_j}\varepsilon_\pm\cdot\vec{r}\ket{\kappa_i}=\frac{|R|}{6}\left[ \sum_p e^{ip\left(\kappa_i-\kappa_j\pm \pi/3\right)}\right]\times \nonumber \\
\left(1-2D(\cos(\kappa_i)-\cos(\kappa_i\pm \pi/3))\right).
\end{eqnarray}
If we collect all the terms outside of the summation into $C(\kappa_i)$, the dipole element between molecular levels can be written as
\begin{equation}
d(\kappa_j,\kappa_i) =\bra{\kappa_j}\vec{\varepsilon}_{\pm}\cdot\vec{r}\ket{\kappa_i}=C(\kappa_i)\delta(\kappa_i-\kappa_j\pm\frac{\pi}{3}),
\label{momconv}
\end{equation}
to give the selection rule for optical transitions - light only couples the molecular states $\ket{\kappa_i}$ and $\ket{\kappa_f}$ that differ by $\pm\pi/3$.

  \section{Electronic structure of charged artificial benzene ring}

We now focus on the charged artificial benzene ring. Removing(adding) an electron from the half-filled ABR creates a hole(electron) in a charge neutral ABR. The hole can be thought as moving in the presence of $N_e=5$ electrons with the total spin projections of $S_z=\{5/2,3/2,1/2\}$. We now proceed to derive analytically the energy spectrum of the hole dressed by the electronic cloud with different total spin $S$ for very strong interactions, $U=\infty$, such that double electronic occupancy is not allowed.  In this strongly interacting regime, it is convenient to work in the real-space basis of the ABR. 

\subsection{Strong interactions and emergence of an artificial gauge in the spectrum of a hole }
\subsubsection{Hole in a spin polarized electronic state $\mathbf{S_z=5/2}$}
In the maximally polarized subspace, $S_z=5/2$, the $5$ spin-polarized electrons are distributed on 6 dots, leaving a hole in the $m$-th dot:
\begin{equation}
\ket{h_m}=c_{m\downarrow}\prod_i^6 c^+_{i\downarrow}\ket{0}.
\label{52hole}
\end{equation}
Just like the QQD\citep{ozfidan_ssc2013} or the TQD\citep{korkusinski_gimenez_prb2007}, the Hamiltonian for a hole in a spin polarized, half filled system and that of a single electron given in \eq{Hammat} are the same except for the change of sign of $t$, resulting in the same single particle energy spectrum as depicted in \Fig{fig2} shifted in energy by $5E+4V$ due to the presence of the electrons residing on 5 dots. 

\subsubsection{ Hole in the presence of a minority spin $\mathbf{S_z=3/2}$ }
The $S_z=3/2$ configurations contain a minority spin  obtained by flipping the spin of one electron in each hole-state of the $S_z=5/2$ subspace such that
\begin{equation}
\ket{j,{h_m}}=c^+_{j\uparrow}c_{j\downarrow}\ket{h_m},
\end{equation}
where $\ket{h_m}$ is the $S_z=5/2$ hole state defined in \eq{52hole} and $j$ is the index of the minority electron with flipped spin. We can take the Fourier transform of the minority spin state, $\ket{j,{h_m}}$, that tunnels among the 5 filled states of a quasi-hole state $\ket{h_m}$, acquiring a phase of $\xi$ every time it tunnels. Upon this rotation, we obtain the states $\ket{\xi,h_m}=\sum_j e^{ij\xi}\ket{j,{h_m}}$ where allowed values of wavevector $\xi$ are $2\pi/5\{0,1,2,3,4\}$. The Hamiltonian becomes block diagonal in $\xi$, such that each block is made out of configurations with the hole in one of the 6 dots, sensing the minority spin chirality $\xi$. Each one of the five 6x6 block is equivalent to the single-particle Hamiltonian, \eq{Hammat}, with the artificial gauge field $e^{i\xi}$ emerging for the hole tunnelling between quantum dots one and six within the chirality space of $\xi$, resulting in a net phase accumulated on the tunneling matrix element $\tau$ , $\tau=t e^{i\xi}$. The Hamiltonians can be analytically diagonalize by realizing that the hole acquires a phase of $\phi_\xi=\xi/6$ every time it tunnels from one dot to another within each minority-spin-chirality-$\xi$. Then the eigenvectors, $\ket{\alpha,\xi}$, are obtained as: 
\begin{equation}
\ket{\alpha,\xi}=\frac{1}{\sqrt{6}}\sum_{m}^6 e^{im\phi_\xi}e^{im\alpha }\ket{\xi,h_m}
\end{equation}
and $E_{\alpha,\xi}=5E+2t\text{cos}(\alpha+\phi_\xi)$, respectively for $\alpha=2\pi/6\{0,1,2,3,4,5\}$. We see that the wavevector of the hole is a combination of the bare wavevector $\alpha$ and the wave vector $\phi_\xi$ of the minority spin current of the background electrons.

\begin{figure}[th]
\begin{center}
\includegraphics[width=0.7\linewidth]{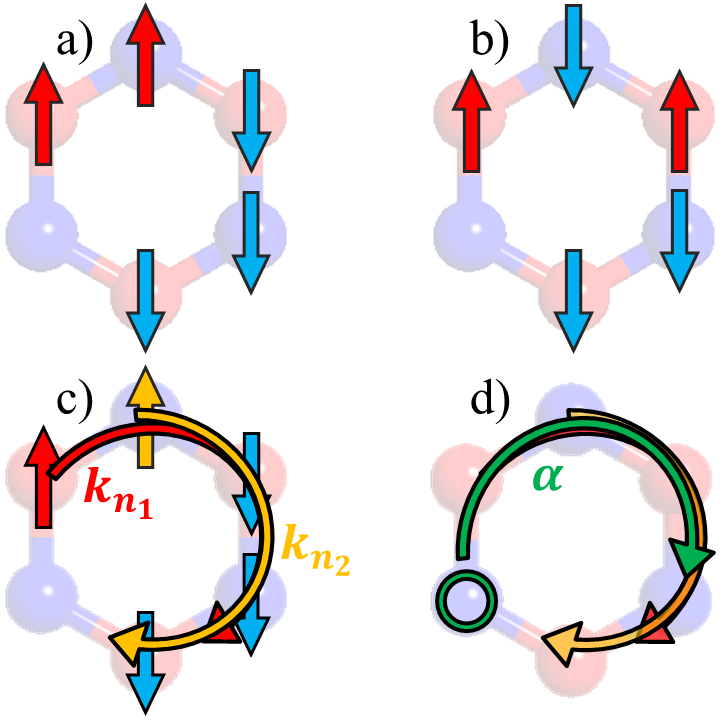}
\end{center}
\caption{(a) and (b): Permutation configurations for two minority spin, $S_z=1/2$, in 5 electron ABR. For a quasi-hole at the lower left dot, (a) two adjacent minority electrons together (b) two minority electrons separated by a majority spin. (c) Figure depicting a spin-current state with the beating of minority spin phases $k_{n_1}$ and $k_{n_2}$ (d)Hole tunnelling under the influence of spin-current-chirality}
\label{fig:perm}
\end{figure}

\subsubsection{Hole in the spin depolarised $\mathbf{S_z=1/2}$ state}
The $S_z=1/2$ subspace requires flipping the spin of two electrons in every $S_z=5/2$ quasi-hole configuration.  
This can be done in two ways. We can (A) flip the spins of two adjacent electrons or (B) two electrons that are separated.
For example, starting with the $\ket{h_6}$ state as defined in \eq{52hole}, the spin of the two electrons can be flipped to give 
$\ket{A_1^6}=\ket{\cu{1}\cu{2}\cd{3}\cd{4}\cd{5}}$ and $\ket{B_1^6}=\ket{\cu{1}\cd{2}\cu{3}\cd{4}\cd{5}}$ as depicted in \Fig{fig:perm}(a) and (b) respectively, where the superscript 6 represents the position of the quasi-hole and subscript 1 represents the configuration-index. 
Applying a permutation operator, $\hat{P}$, which moves all electrons to the right by one dot\citep{caspers_iske_physe1989}, $\hat{P}\ket{A_1^6}=\hat{P}\ket{\cu{1}\cu{2}\cd{3}\cd{4}\cd{5}}=\ket{\cd{1}\cu{2}\cu{3}\cd{4}\cd{5}}=\ket{A_2^6}$, we obtain 4 other permutations of $\ket{A_1^6}$ and $\ket{B_1^6}$. The Hamiltonian separates into blocks of $\ket{A}$ and $\ket{B}$ configurations. 
Configurations $A$ ($B$) correspond to a pair of minority spin electrons moving on the ring of 5 electrons.  
 Just as in the $S_z=3/2$ case, for a given hole, state we can take the Fourier transform of the five $A^h (B^h)$ minority spin pair configurations to obtain the states differentiated by the phase $\varphi=2\pi/5\{0,1,2,3,4\}$, and generate 6x6 blocks for each $\varphi$ representing a quasi-hole tunnelling under the influence of the artificial gauge field in the form of \eq{Hammat} with $\tau=te^{i\varphi}$. Upon rotating each block, one finds that the $A$ and $B$ subspaces are degenerate. 

Although the permutation operator provides a convenient way to describe the dressed quasi-hole states, the states obtained by this method are not eigenvectors of the total spin operator $\hat{S^2}$.

\subsubsection{ Total spin classification of $\mathbf{S_z=1/2}$ hole states }

In order to obtain the eigenstates of the total spin operator, we introduce the spin current operator $\hat{J}_n$. The spin-current operator takes an electron from an $S_z=5/2$ quasi-hole state $\ket{h_m}$, flips its spin and moves it among the occupied dots, adding a phase of $e^{ik_{n}}$ each time it tunnels such that:

\begin{equation}
\hat{J}_n=\sum_j e^{ik_{n}j}c^+_{j\uparrow}c_{j\downarrow},\hspace{0.1in}k_n=\frac{2\pi}{5}n,\hspace{0.1in}n=\{1,2,3,4,5\}.
\end{equation}
For $S_z=1/2$ subspace, one needs to apply spin-current operators twice,
\neqn{\hat{J}_{n_2}\hat{J}_{n_1}=\sum_j\sum_l e^{i(jk_{n_1}+lk_{n_2})}\cd{l}\au{l}\cd{j}\au{j},}
on to the spin polarized $\ket{h_m}$ state. In this process there appears 25 $(j,l)$ pairs of minority spin electrons at sites $j$ and $l$ and 25 current states $(k_{n_1},k_{n_2})$. Not all of these configurations have nonzero amplitudes. The spin-current states, 
$\{k_{n_1},k_{n_2}\}$, and the $(j,l)$-pairs created from the beating of the two phases $k_{n_1}$ and $k_{n_2}$ carried by the minority spins (\Fig{fig:perm}(c)) are  orthogonalized, removing duplicates that emerge due to indistinguishable nature of electrons. Out of the 25 $(j,l)$ pairs, 5 do no exist since $\cd{l}\au{l}\cd{j}\au{j}\ket{h_m}=0$, and the remaining 20 are made out of duplicates since $\cd{l}\au{l}\cd{j}\au{j}=\cd{j}\au{j}\cd{l}\au{l}$, leaving only 10 distinct $(j,l)$-pairs. However, removing the 5 non-existent $(j,l)$ pairs destroys the orthogonality of the spin-current states which require re-orthogonalization. 

Upon closer examination of the spin-current states, one can see that $\{k_{n_1},k_{n_2}\}=\{k_{n_2},k_{n_1}\}$, which automatically removes 10 out of these 25 spin-current states leaving 15 to work with. Though at first glance these 15 spin-current states seem to be distinct, we expect to have only 10 states at the end of this process, and upon re-orthogonalization, we will see that there are 5 duplicates, leaving 10 distinct spin-current states. We start by grouping the spin-current states $\{k_{n_1}, k_{n_2}\}$ according to their total spin current, $k_{tot}=k_{n_1}+k_{n_2}=\frac{2\pi}{5}\{1,2,3,4,5\}$  (in units of $2\pi/5$):
\vspace{0.15in}
\begin{center}
\begin{tabular}{|c|c|c|c|c|}
 $k_{tot} = 1$ &  $k_{tot} = 2$ &  $k_{tot} = 3$ &  $k_{tot} = 4$ &  $k_{tot} = 5$ \\
\hline
$\{k_1,k_5\}$ & $\{k_2,k_5\}$ & $\{k_3,k_5\}$ & $\{k_4,k_5\}$ & $\{k_5,k_5\}$ \\
$\{k_3,k_3\}$ & $\{k_1,k_1\}$ & $\{k_4,k_4\}$ & $\{k_2,k_2\}$ & $\{k_1,k_4\}$ \\
$\{k_2,k_4\}$ & $\{k_3,k_4\}$ & $\{k_1,k_2\}$ & $\{k_1,k_3\}$ & $\{k_2,k_3\}$ \\
\end{tabular}
\end{center}
Above states are orthogonal to one another if they belong to different total-spin-current subspaces, yet within each subspace they are not. Acting with the $\hat{S}^2$ operator, one can see that all $\{k_{n_1},k_5\}$ states belong to the $S=3/2$ space except for $\{k_5,k_5\}$ which is the only state with $S=5/2$. When the $\hat{S}^2$ operator acts on the remainder of the states, we see that they are not eigenfunctions of the total S operator. These states, with no definite spin, are orthogonalized using the Gram-Schmidt method revealing that both of the undefined total spin states within a given subspace are actually one another's duplicate, resolving the problem of 5 excess spin-current states. 

\begin{figure}[h]
\begin{center}
\includegraphics[width=0.98\linewidth]{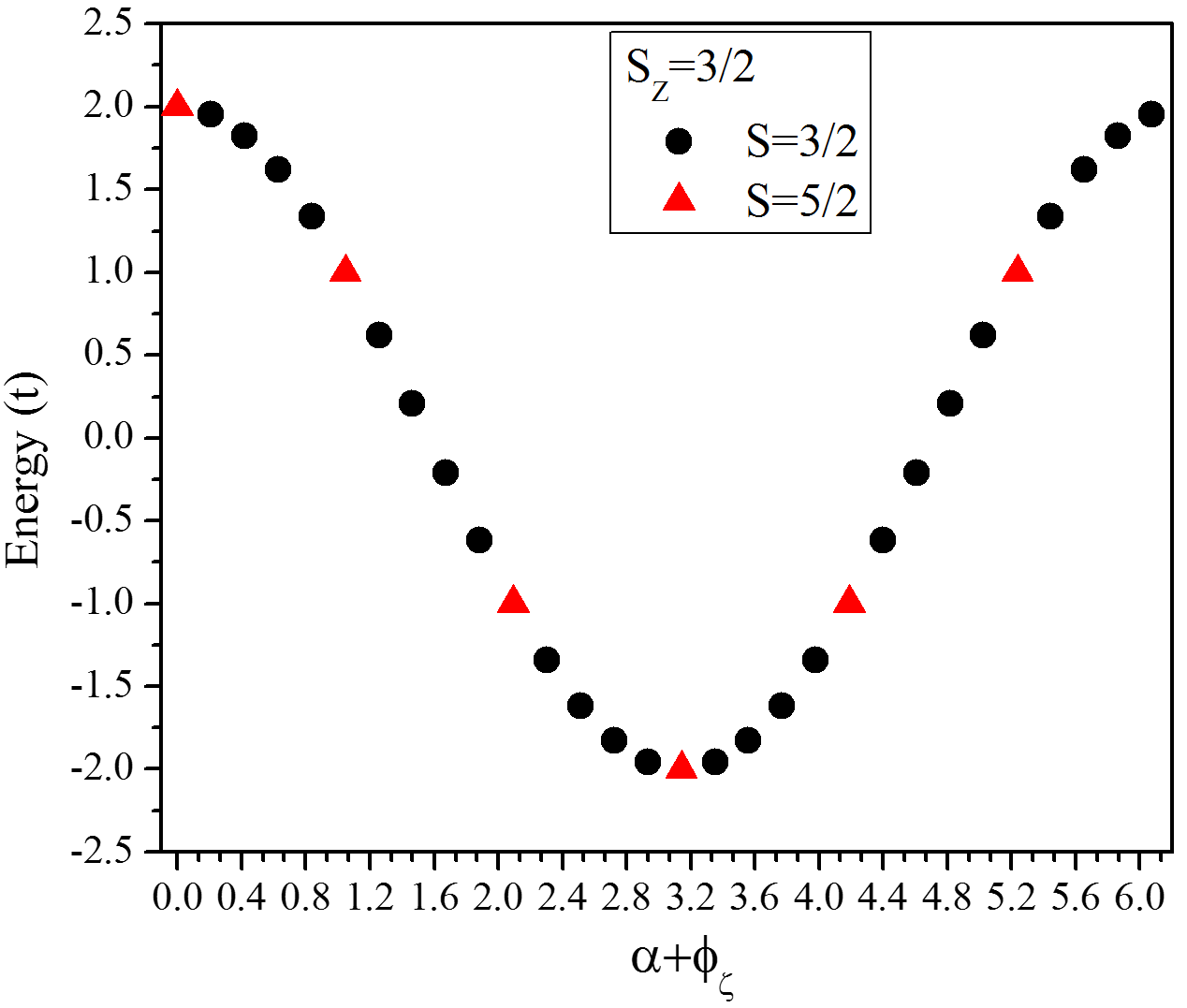}
\end{center}
\caption{Figure shows the energies of the $S=5/2$ and $S=3/2$ states as a function of total wavevector $\alpha+\phi$ or total phase. Each energy level shown above is degenerate with energy levels of  $S=1/2$ states as discussed in the text.U$\rightarrow \infty$ limit, V=0. }
\label{fig5}
\end{figure}
Then, from each $S_z=5/2$ quasi-hole state, applying the spin-current operator twice, one arrives at 10 total-spin current states, with 5 distinct $k_{tot}$. Just as in $S_z=3/2$ case, we can now divide the Hamiltonian into 10 subspaces, each belonging to a different total-spin-current, total-spin, $\{k_{tot},S\}$ pair. Again, the Hamiltonians of each $\{k_{tot},S\}$-subspace, made out of six vectors $\ket{k_{tot},S, h_m}$ for each hole position, are similar to that of a single-electron Hamiltonian (\eq{Hammat}) with an additional $5E+4V$ energy on the diagonals and the tunnelling matrix element between dots 1 and 6 is modified by the phase the quasi-hole acquires when dressed by the spin-current, $\tau=te^{ik_{tot}}$ (\Fig{fig:perm}(d)). From the phase, one can deduce that the energy spectrum is doubly degenerate since there is a $S=1/2$ and a $S=3/2\text{ or }5/2$ subspace for each $k_{tot}$. The following eigenfunctions 
 \begin{equation}
 \ket{\chi_{k_{tot},S}^{\alpha}}=\frac{1}{6}\sum_{m}^6 e^{i\cdot m\cdot \phi_{k_{tot}}}e^{i\cdot m\cdot \alpha}\ket{k_{tot},S,{h_m}},
 \label{eig}
 \end{equation}
in which the hole gains one-sixth of the total phase, $\phi_{k_{tot}}=(k_{tot})/6$, every time it tunnels from one dot to another, diagonalize the Hamiltonian (\eq{Hammat}) with the phase $\tau=te^{ik_{tot}}$, where $\alpha=\frac{2\pi}{6}\{0,1,2,3,4,5\}$. 
\Fig{fig5} depicts the allowed energy levels. As derived above, the $S_z=1/2$ subspace contains all possible total spin states, $S=\{5/2, 3/2, 1/2\}$, and every $S=1/2$ state has a degenerate, higher spin pair. Then the spectrum for the $S_z=3/2$ subspace, which covers both the $S=3/2$ and the $S=5/2$ total spin states, includes all allowed energy levels. We see that the hole moving in the space of polarized spins ($S_z=S=5/2$) is restricted to only 5 energy levels. Whereas when we introduce a minority spin, its chirality acts as an additional wavevector and allows the hole to be on more than 5 different states.  

\subsection{Quasi-hole in a weakly interacting system}
Let us now study a weakly interacting system, $U<<t$, where the electronic properties are determined primarily by the kinetic energy, with interactions acting as a perturbation. Thus working in the Fourier-space, where the kinetic energy has already been diagonalized simplifies our discussion considerably. 

\begin{figure}[h]
\begin{center}
\includegraphics[width=0.98\linewidth]{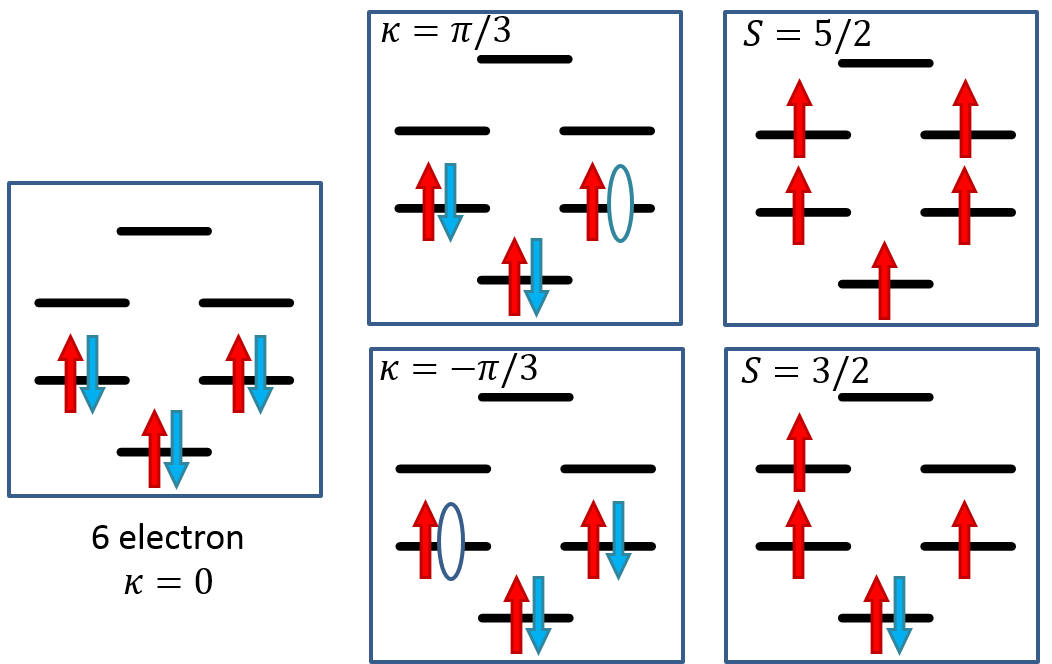}
\end{center}
\caption{The ground state configuration in weakly interacting regime plotted in Fourier space. For $U=0$, all electrons will occupy the lowest kinetic energy levels. For 6 electrons that translates to double occupancy. When an electron is removed there is a degenerate ground state. Flipping a spin from either one of them will generate a $S=3/2$ ground state. Fully polarized electrons will occupy the lowest levels with single occupancy.}
\label{fig4}
\end{figure}

In the non-interacting limit, $U\rightarrow0$, we place electrons on the single particle levels while satisfying the Pauli exclusion principle. For a half-filled ABR the reference state, a Fermi sea, illustrated in \Fig{fig4}(a), is our starting configuration. Removing an electron creates a hole below the Fermi level. There are two degenerate configurations for the hole as depicted in \Fig{fig4}(b). Since $\kappa=\pm\pi/3$ levels are degenerate, creating a hole in either one costs the same energy. From each of these degenerate states with spin $S=1/2$, one can create the lowest energy $S=3/2$ configurations with energy $E_{3/2}^{\text{GS}}=-5t$ as depicted in \Fig{fig4}(c, S=3/2). Just like the $S=1/2$ ground state, the $S=3/2$ ground state is also degenerate due to the degeneracy of the $\kappa=\pm2\pi/3$ levels. Finally, the lowest energy $S=5/2$ state is obtained by placing a single electron on each one of the lowest 5 levels, resulting in $E_{5/2}^{\text{GS}}=-2t$ and $\kappa_{5/2}^{\text{GS}}=0$ as shown in \Fig{fig4}(c, S=5/2). Unlike its lower total spin counterparts, this ground state is non-degenerate since there is only one way of placing five spin-polarized electrons on to the lowest five levels. Then, in the weakly interacting regime, the ground state of the charged ABR has a unique total spin but the degeneracy arises from the degeneracy of the orbitals. This is to be contrasted with the strongly interacting regime where the two unique total spin states, $S=1/2$ and $S=5/2$, are degenerate. The difference in the nature of the ground state in the two limits implies level crossing as a function of the interaction strength.

\section{Numerical results for intermediate interaction strength  $0<U<\infty$}
\begin{figure}[t]
\begin{center}
\includegraphics[width=0.98\linewidth]{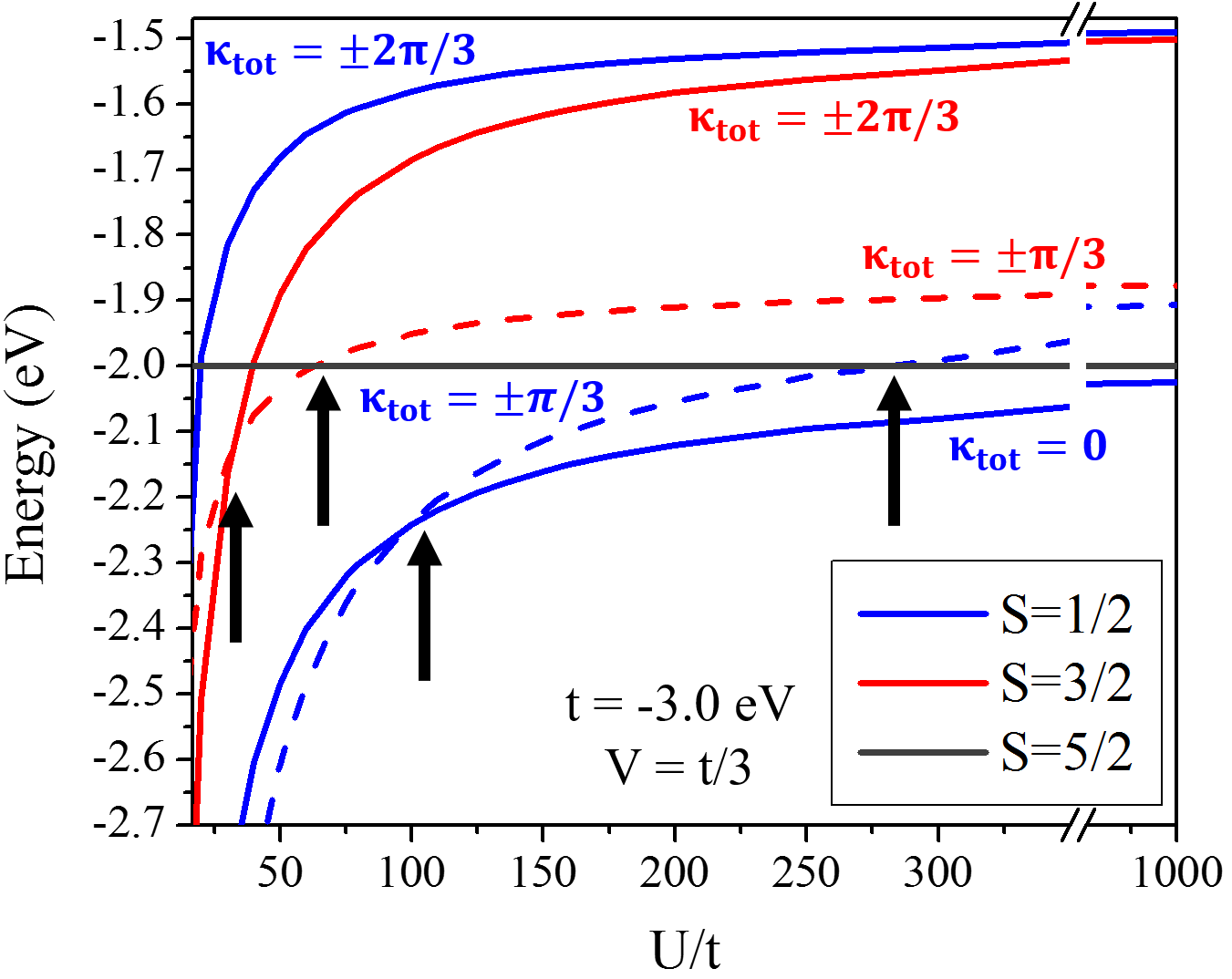}
\end{center}
\caption{Transitions in the lowest energy levels with increasing interaction strength U. The black arrows highlight the transition points. Blue, red and gray colors correspond to $S=1/2$, $S=3/2$ and $S=5/2$ states while solid and dashed lines distinguish the $\kappa_{tot}$ of these subspaces.}
\label{abstrans}
\end{figure}

For finite $t,U,V$ we diagonalize numerically the Hamiltonian matrix in the space of 5 electron configurations in Fourier space.
 The evolution of the low energy levels with increasing $U$ is shown in \Fig{abstrans}. Starting from the weakly interacting regime, the first level crossing as we turn on the interactions is found in the excited $S=3/2$ subspace where the wavevector of the lowest energy state changes from $\kappa=\pm 2\pi/3$ to $\kappa=\pm\pi/3$. Next, the crossing between the degenerate $\{S=3/2, \kappa_{tot}=\pm\pi/3\}$ states and the lowest $\{S=5/2, \kappa_{tot}=0\}$ state changes the total spin order of excited states. As we keep increasing the interaction strength, the ground state of the ABR undergoes a transition from the $S=1/2$, degenerate in momentum $\kappa=\pm\pi/3$ states to a non degenerate $S=1/2$, $\kappa=0$ state. As $U$ grows, the energy of the $S=1/2$ ground state approaches the $\{S=5/2, \kappa_{tot}=0\}$ state, eventually becoming degenerate as predicted and derived in the previous section.

\section{Absorption spectrum of a charged artificial benzene ring}

Here we will analyse how  the  electron-electron interaction driven transitions in the ground and excited states 
can be detected by optical spectroscopy. The transition from a degenerate, $\kappa=\pm\pi/3$, ground state to a non degenerate $\kappa=0$ angular momentum ground state can be captured in the absorption spectrum using the selection rules on angular momentum. We have already derived the selection rules for angular momentum and  photons conserve the total spin of the system in the absence of spin-orbit interaction.

\subsection{Absorption spectrum of a weakly interacting charged artificial benzene ring}

\begin{figure}[t]
\begin{center}
\includegraphics[width=0.98\linewidth]{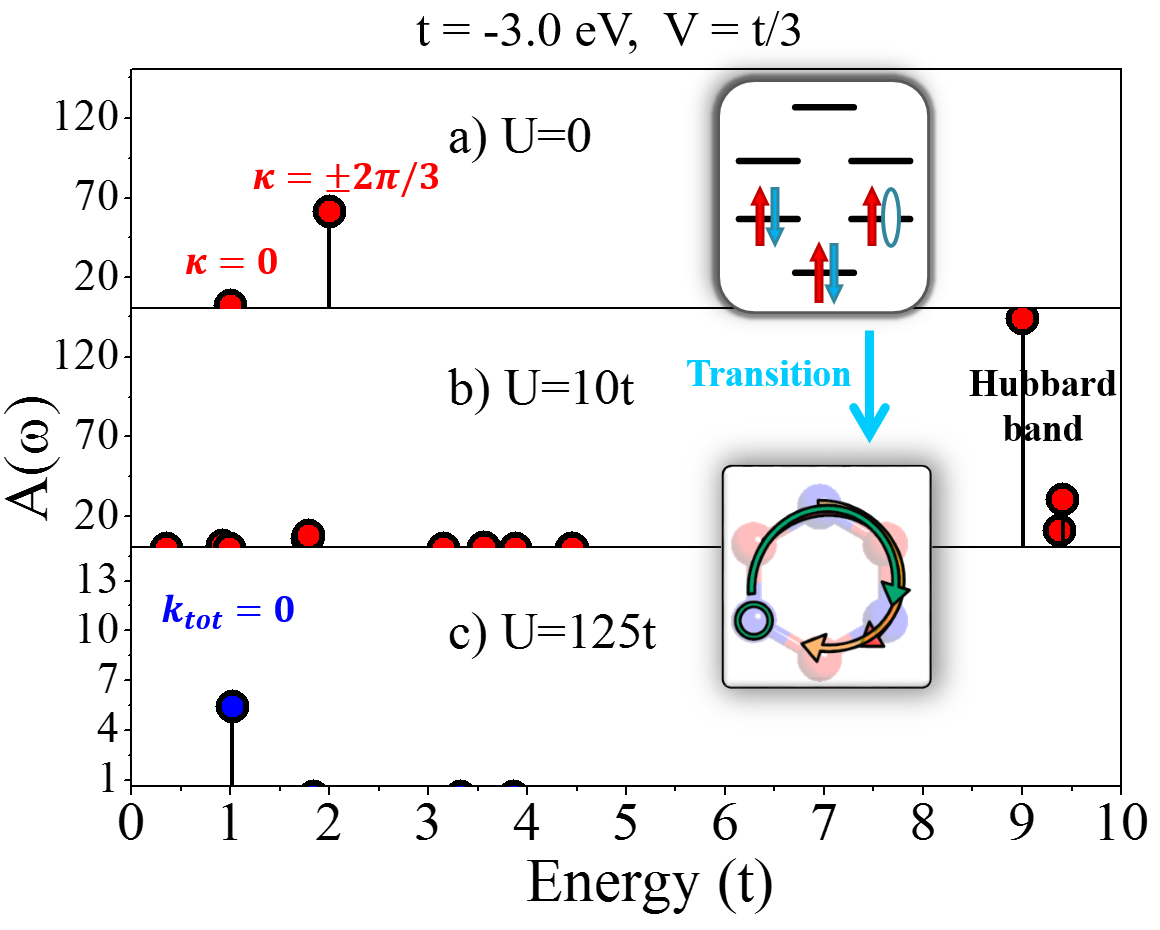}
\end{center}
\caption{Absoprtion spectrum of charged ABR with changing U/t. V is kept fixed at $V=t/3$. (a)  Absoprtion spectrum for noninteracting ABR ( U=0 ). (b)  Absoprtion spectrum for interacting ABR ( U=10t ),  showing the splitting of transitions at $E=2 t$into many lines, appearance of new low energy transitions and emergence of transitions at $E~U$, th e excited Hubbard band. (c) Optical spectrum at $U>>t$ at $E= 1 t $. The inset highlights the fact that as the interaction strength increases molecular levels becomes highly correlated and a real space representation is necessary for a better understanding of the problem.}
\label{abs5e}
\end{figure}

In the non-interacting limit, $U=0$, the absorption is solely dictated by the single particle level selection rules. Starting with either one of the two $\kappa_{tot}=\pm\pi/3$, $S=1/2$ states ($\kappa_{tot}=-\pi/3$ shown as inset in \Fig{abs5e}), an electron from the $\kappa=0$ level can be excited to the singly occupied $\kappa=\mp\pi/3$ level via a photon with energy $\omega=t$. For the cost of $2t$, either one of the electrons in the doubly occupied $\kappa=\pm\pi/3$ can be moved to $\kappa=\pm2\pi/3$. In the non-interacting regime, other transitions are not allowed due to optical selection rules, leading to two absorption lines at $\omega=t$ and $\omega=2t$ shown in \Fig{abs5e}(a). 

\begin{figure}[t]
\begin{center}
\includegraphics[width=0.98\linewidth]{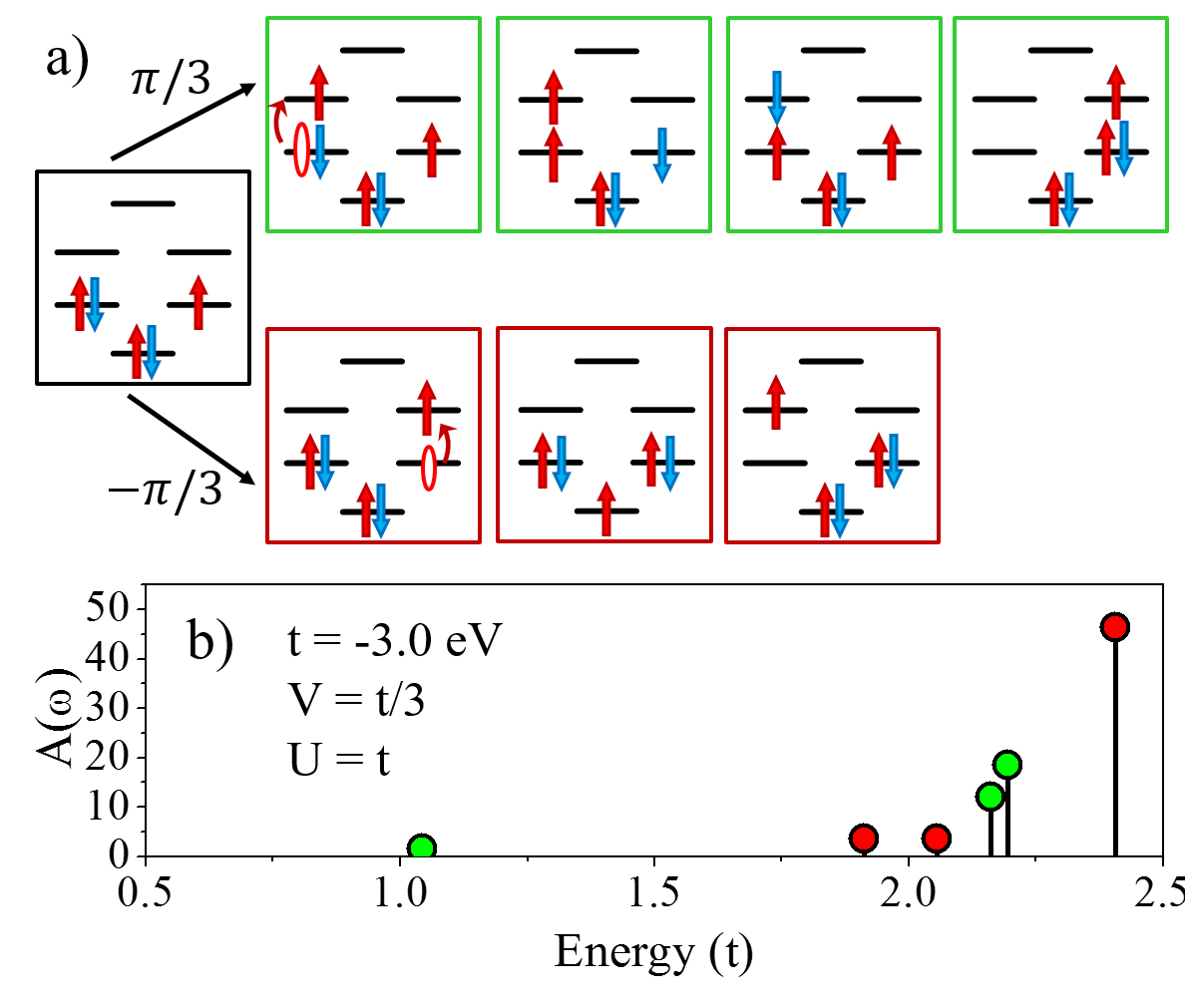}
\end{center}
\caption{The optically allowed configurations(a) and the absorption spectrum from the weakly interacting, degenerate ground state(b). The configurations responsible for the low energy peaks in the spectrum are shown in green and red boxes. Starting with the ground state with $\kappa_{tot}=\pi/3$, the peaks in the absorption spectrum that correspond to $\kappa_{tot}=2\pi/3$ (in green) and the $\kappa_{tot}=0$ (in red) excitations are identified and separated. $t=-3.0$eV, $U=t$, $V=t/3$.}
\label{absdet}
\end{figure}

When the interactions are turned on, multiple configurations contribute to the absorption spectrum as shown in \Fig{absdet}(a) for $U=t$. Starting with the state with $S=1/2$, $\kappa_{tot}=\pi/3$, a photon can only couple this state to $\kappa_{tot}=2\pi/3,0$ states. Due to interactions, one needs to consider correlated states within each total wavevector subspace. 

Let us concentrate on the configurations with the greatest contribution to the absorption spectrum within each of the $\kappa_{tot}$ subspace. For the weakly interacting regime, these are the lowest kinetic energy configurations.  There are four configurations with kinetic energy $-5t$ in the $\kappa_{tot}=2\pi/3$ subspace as depicted in green boxes in \Fig{absdet}(a). They all have two electrons in the lowest, $\kappa=0$, molecular state and the remaining three electrons are distributed on the degenerate levels. 
Absorption transitions from the $\kappa_{tot}=\pi/3$ ground state to a superposition of the $\kappa_{tot}=2\pi/3$ configurations are plotted in green in \Fig{absdet}(b). Although some of the configurations shown in the boxes are not directly optically accessible from the ground state, they  contribute to the correlated states and hence acquire finite oscillator strength. Since the cost of moving an electron across the degenerate levels is $2t$, the green peaks in the absorption spectrum are around $2t$, shifted left and right due to correlations.  

The same analysis can be done for the $\kappa_{tot}=0$ subspace. The lowest energy configurations within this subspace are depicted in red boxes in \Fig{absdet}. The lowest kinetic energy, $t$, excitation, in which an electron from the $\kappa=0$ molecular level is moved up to the $\kappa=-\pi/3$ molecular level, interacts with higher energy, $2t$, excitations through off diagonal terms. The red absorption peak around $E=|t|$ corresponds to a state  mainly composed of the low energy configuration. Since the cost of moving an electron across the degenerate levels costs $2t$, the other two peaks that are mainly composed of the higher energy $\kappa_{tot}=0$ states, are around $E=2|t|$. 

Let us return to the absorption spectrum for the interacting ABR shown in \Fig{abs5e}b. In the calculated absorption spectrum for the interacting 
ABR ($U=10t$), the two peaks at $t$ and $2t$ that are characteristic of the non-interacting system, split into many lines. At higher energies, $E\sim U$, there appears a new band of transitions to the first Hubbard band. These excited states correspond to creation of ``holons'' (empty sites)  and ``doublons'' (doubly occupied states)\cite{dagotto_prl2008}.

As we increase $U/t$ further, the $S=1/2$ ground state of the charged ABR changes from the degenerate $ \kappa_{tot}=\pi/3$ states to the non-degenerate $\kappa_{tot}=0$ state which approaches the spin polarised $S=5/2$ state as shown in \Fig{abstrans}. The optical transitions to the first Hubbard band move to higher energies and the low energy absorption spectrum from the $S=1/2$, $\kappa_{tot}=0$, ground state simplifies to 
an absorption peak at $E=t$ as shown in \Fig{abs5e}. We can understand the absorption spectrum for $U\rightarrow\infty$ and its relationship with the quasi-hole energy spectrum (\Fig{fig5}) determined by emergence of the artificial gauge, by deriving the selection rule for the quasi-hole position, $\Delta m=(m_1-m_2)=0,\pm 1$ where $\pm 1$ applies if the hole is tunnelling between the 1st and the 6th dot. Next, upon evaluation of the dipole matrix elements between total-spin-current quasi-hole states, 
\begin{equation}
\bra{k_{tot}^1,S_1,h_{m_1}}\varepsilon_{\pm}\cdot\vec{r}\ket{k_{tot}^2,S_2,h_{m_2}}=C_{m_1}^{m_2}\delta(k_{tot}^1-k_{tot}^2),
\end{equation}
the conservation rule on $k_{tot}$ is obtained. In the equation above, $C_{m_1}^{m_2}$ is the dipole strength that depends on the positions of the hole. Conservation of $k_{tot}$ means that, only absorption within total-spin-current subspaces are allowed. Now, if the dipole matrix elements between total-spin-current eigenvectors ,  $\ket{\chi_{k_{tot},S}^{\alpha}}$, are calculated remaining within the same total-spin-current subspace and using the selection rules on the quasi-hole position, we get,
\begin{eqnarray}
\bra{\chi_{k_{tot},S}^{\alpha_1}}&\varepsilon_\pm\cdot\vec{r}&\ket{\chi_{k_{tot},S}^{\alpha_2}}=\frac{1}{6}\sum_{m_1}e^{i\cdot m_1(\alpha_2-\alpha_1\pm 1))} \nonumber \\
&\text{x}&\left[1+Ce^{ik_{tot}\delta(m_1,1)}+C^*e^{-ik_{tot}\delta(m_1,6)}\right]. \nonumber \\
\end{eqnarray}
Even though  there seems to be a selection rule, $\alpha_2=\alpha_1\pm1$ on the total-spin-current eigenvector index, it is destroyed by the fact that there is hole-position dependence in $\delta(m_1,1), \delta(m_1,6)$. However, for the ground state where $k_{tot}=0, \alpha=0$, the selection rule is restored and only $\alpha_2=\alpha_1\pm1$ transitions are allowed. Since the energies of $\alpha=1$ and $\alpha=5$ states are degenerate, we find a single peak in the absorption spectrum obtained numerically  for U$=125t$  in \Fig{abs5e}(c).

\section{Conclusion}
We presented here a theory of a charged artificial benzene ring (ABR)  described by the extended Hubbard model.  We derived an exact expression for the energy spectrum of the quasi-hole in a half-filled ABR expressed in terms of the emergent topological gauge field in the limit of strong interactions. Using exact diagonalization techniques, we have described the evolution and transitions in the ground state spin and momentum as a function of the interaction strength.  The evolution of the ground and excited states in the optical absorption spectrum was predicted and analyzed.  It is hoped that the obtained results will stimulate research on artifical benzene rings fabricated using semiconductor quantum wires.  

Acknowledgment: I.O., M.D, and P.H. acknowledge support of NSERC.

\end{document}